\documentclass[%
 reprint,
 twocolumn,
 superscriptaddress,
 prb,
]{revtex4-2} 

\usepackage[T1]{fontenc}    
\usepackage[utf8]{inputenc} 
\usepackage[english]{babel} 

\usepackage{amsmath}
\usepackage{amssymb}
\usepackage{mathrsfs}   
\usepackage{mathtools}  
\usepackage{dsfont}     

\usepackage{microtype}  

\usepackage{physics}
\usepackage{siunitx}

\usepackage{epsfig}
\usepackage{epstopdf}
\usepackage{graphicx}
\graphicspath{{fig/}}
\usepackage{dcolumn}
\usepackage{bm}
\usepackage{xcolor}
\usepackage{changes}

\usepackage[unicode]{hyperref}
\usepackage[all]{hypcap}
\usepackage[capitalize]{cleveref}   
\usepackage{arcs}

\begin{document}
\author{A.\ \"Ostlin}
\affiliation{Theoretical Physics III, Center for Electronic Correlations and Magnetism, Institute of Physics, University of Augsburg, 86135 Augsburg, Germany}
\affiliation{Augsburg Center for Innovative Technologies, University of Augsburg, 86135 Augsburg, Germany}
\author{I.\ Titvinidze}
\email{irakli.titvinidze@uni-a.de}
\affiliation{Theoretical Physics III, Center for Electronic Correlations and Magnetism, Institute of Physics, University of Augsburg, 86135 Augsburg, Germany}
\author{M.\ Kollar}
\affiliation{Theoretical Physics III, Center for Electronic Correlations and Magnetism, Institute of Physics, University of Augsburg, 86135 Augsburg, Germany}
\author{D.\ Jones}
\affiliation{Theoretical Physics III, Center for Electronic Correlations and Magnetism, Institute of Physics, University of Augsburg, 86135 Augsburg, Germany}
\author{L.\ Chioncel}
\affiliation{Theoretical Physics III, Center for Electronic Correlations and Magnetism, Institute of Physics, University of Augsburg, 86135 Augsburg, Germany}
\affiliation{Augsburg Center for Innovative Technologies, University of Augsburg, 86135 Augsburg, Germany}

\date{\today}

\title{
Analytical results of Typical Medium Theory \\ 
for the Bethe lattice with Cauchy disorder}

\begin{abstract}
We present a combined numerical and  analytically tractable realization of the typical medium theory (TMT) for Anderson localization on the infinite-connectivity Bethe lattice with Cauchy-distributed on-site disorder. 
Exploiting the special properties of the Cauchy distribution and the Bethe lattice, we derive an analytical expression for the typical density of states (TDOS) and obtain a simplified TMT self-consistency scheme. 
We show that the TDOS at the band center decreases linearly with disorder strength and vanishes at the critical disorder $W_c=D/2$. The resulting $\omega$--$W$ phase diagram reveals the continuous collapse of the mobility edge 
and the disappearance of extended states.
\end{abstract}

\maketitle

\section{Introduction}
\label{sec:introduction}

Disorder is an inherent feature of condensed matter systems and plays a decisive role in determining their electronic properties. 
While weak disorder generally leads to diffusive transport, sufficiently strong randomness can suppress diffusion through quantum interference, resulting in the Anderson localization transition~\cite{ande.58,ve.ki.68,ab.th.73,el.kr.74, ab.an.79,le.ra.85,ev.mi.08}. This disorder-driven metal-insulator transition is particularly remarkable because it occurs without any accompanying symmetry breaking~\cite{ab.th.73,ab.an.79,wegn.79, mi.fe.94,kr.ma.93} and therefore represents a fundamentally different mechanism from conventional interaction-driven phase transitions~\cite{im.fu.98}. Understanding the nature of Anderson localization and the transition between conducting and insulating regimes remains one of the central problems in condensed matter physics~\cite{kr.kr.94, be.ki.94,ev.mi.08}.

The electronic states of a disordered system can be classified according to their spatial properties as extended, localized, or critical~\cite{kr.ma.93,be.ki.94,mirl.00,pa.pa.20}. Extended states support electronic transport, whereas localized states are confined in space and do not contribute to dc conduction. Critical states emerge at the transition between these two regimes and exhibit properties distinct from both extended and localized states. Although Anderson localization has been studied extensively for several decades~\cite{el.kr.74,be.ki.94,ev.mi.08}, the characterization of critical states and the detailed properties of the localization transition remain active areas of research~\cite{pa.pa.20,ga.ma.22}.

The conventional single-particle Anderson problem describes electrons moving on a lattice in the presence of quenched random potentials. In recent years, related concepts have also been extended to interacting many-body systems, where localization in Fock space leads to the phenomenon of many-body localization (MBL)~\cite{ba.al.06,na.hu.15,ab.al.19,si.le.25}. In this context, random regular graphs and Bethe lattices have played an important role because they provide a mean-field-like setting in which localization phenomena can be investigated while neglecting short loops and spatial correlations~\cite{lo.wo.87,al.ge.97,lu.sc.13,pa.pa.20}. These studies emphasize the importance of understanding localization and critical behavior in high-dimensional and effectively infinite-dimensional systems~\cite{pa.pa.20}.

A particularly useful approach to the Anderson localization problem is based on effective medium theories, where the disordered lattice problem is mapped onto an impurity problem embedded in a self-consistently determined medium~\cite{ve.ki.68,le.ra.85,ev.mi.08}. 
We adopt such an approach and consider non-interacting spin-less fermions hopping on a lattice described by the  Anderson-Hamiltonian
\begin{equation}
\label{equ:H_and}
H =\sum_{i} \epsilon_{i}n_{i}  - t \sum_{\langle{i,j}\rangle} (c^{\dagger}_{i} c_{j}^{\phantom\dagger} + h.c.) \,,
\end{equation}
where $c^{\dagger}_{i}$ and $c_{i}$ are  fermion creation and annihilation operators at site $i$, $n_{i}=c^{\dagger}_{i}c_{i}$ is the occupation number, and $t$ denotes the nearest-neighbor hopping amplitude. The disorder is introduced through the random on-site energies $\epsilon_i$, which are distributed according to a given probability distribution.

Several disorder distributions have been considered in the literature, including box, Gaussian, and Cauchy (Lorentzian) distributions~\cite{le.ra.85,be.ki.94,ev.mi.08}. The Cauchy distribution is of particular interest because the average Green's function of the corresponding Anderson model can be determined exactly for arbitrary hopping, as shown by Lloyd~\cite{lloy.69}. Bishop~\cite{bishop.73} later demonstrated that for Cauchy disorder, the coherent potential approximation (CPA)~\cite{sove.67,tayl.67,ve.ki.68,on.to.68,yone.68,ki.ve.70,shib.71} reproduces the exact disorder-averaged solution. Although the Cauchy distribution is somewhat peculiar due to the divergence of its higher moments, it remains of significant theoretical and experimental interest. In particular, it can be related to the kicked rotor model~\cite{gr.pr.84}, which has been realized experimentally using cold atoms~\cite{to.mc.22,ca.sa.22}.

Despite its success in describing averaged spectral properties, the CPA is unable to capture Anderson localization because it neglects spatial fluctuations of the local environment. Consequently, the transition between extended and localized states, as well as the associated critical states, cannot be described within this approach. 
Similar to CPA, the typical medium theory (TMT)~\cite{do.pa.03, dobro.10} maps the lattice problem onto an effective impurity problem embedded in a self-consistent medium. The essential difference is that CPA is based on the arithmetic average of the Green's function, whereas TMT uses the geometric average of the local density of states, defining the typical density of states (TDOS). Since the distribution of the local density of states changes qualitatively near the localization transition~\cite{sc.sc.10,lo.wo.87,dobro.10, ma.ta.15, te.zh.17, by.ho.05, by.ho.10}, the TDOS  vanishes when states become localized and therefore serves as an indicator of Anderson localization.

An analytically solvable realization of TMT would provide a valuable benchmark for understanding the capabilities and limitations of this approach. Motivated by the exact results for the Cauchy-disordered Anderson model within the Lloyd model framework~\cite{lloy.69}, we apply the TMT formalism to this disorder distribution. We demonstrate that, for the Bethe lattice, the TMT self-consistency equations simplify considerably and the disorder average entering the TDOS can be evaluated analytically. This allows us to obtain closed analytical expressions for the TDOS and to investigate the Anderson localization transition in detail.

The paper is organized as follows. We briefly introduce in Sec.~\ref{sec:tmt_formalism}  the TMT formalism and the corresponding self-consistency loop for a general lattice, while Sec.~\ref{sec:tmt_formalism_bethe} discusses its simplification for the Bethe lattice with Cauchy disorder. In Sec.~\ref{sec:results} we analyze the behavior of the TDOS (Sec.~\ref{sec:tdos_results}), then study the evolution of extended and localized spectral weights (Sec.~\ref{sec:exatended_localized_states}).
In Sec.~\ref{sec:analytical_analysis}, we provide an approximate analytical description of the metallic TMT solution, deriving analytical expressions for the typical Green's function and the self-energy in the weak-disorder regime. The paper concludes with a summary of the main results in Sec.~\ref{sec:conclusion}.
The paper also includes four appendices with technical details.

\section{TMT-formalism: Brief overview}
\label{sec:tmt_formalism}

The typical medium theory (TMT) can be viewed as a generalization of the coherent potential approximation (CPA). Similar to the dynamical mean-field theory (DMFT) \cite{ge.ko.96,me.vo.89} and the CPA  \cite{sove.67,tayl.67,ve.ki.68,on.to.68,yone.68,ki.ve.70,shib.71}, the TMT maps the original lattice problem onto an effective single-site impurity embedded in a self-consistently determined medium \cite{do.pa.03,dobro.10,ma.ta.15}. 
The key qunatity of TMT is the TDOS which serves as the order parameter for Anderson localization and is used to construct the self-consistent effective medium.

For a general lattice, the self-consistent TMT procedure consists of the following steps:

\begin{enumerate}

\item
Start from an initial guess for the self-energy. A natural choice is $\Sigma^{W}(\omega)=0$ , where the subscript $W$ indicates presence of disorder with strength $W$.

\item
For a given self-energy $\Sigma^W(\omega)$, the hybridization function is obtained from the Dyson equation
\begin{equation*}
\label{equ:hybridization}
\Delta^W(\omega) = \omega^{+} - \Sigma^W(\omega) - \frac{1}{G_0\!\left(\omega-\Sigma^W(\omega)\right)} \,,
\end{equation*}
where $\omega^{+}=\omega+i\eta$ and $\eta\rightarrow0^{+}$. Here,
\begin{equation*}
\label{equ:local_green}
G_0(\omega) = \int_{-\infty}^{\infty} d\omega' \,
\frac{\rho_0(\omega')} {\omega^{+}-\omega'}
\end{equation*}
is the local Green's function of the non-disorder (clean), non-interacting system, with $\rho_0(\omega)$ denoting its density of states (DOS).

\item
For each value of the local disorder potential $\epsilon_i$, calculate the impurity Green's function
\begin{equation*}
\label{equ:impurity green}
G^W(\omega,\epsilon_i) = \frac{1}{\omega^{+}-\epsilon_i-\Delta^W(\omega)}.
\end{equation*}

\item
Construct the TMT mean-field order parameter, namely the TDOS,
\begin{equation}
\label{equ:avg}
\rho_{\mathrm{typ}}^W(\omega) = \exp\!\left[\int d\epsilon_i\, P^W(\epsilon_i)
\ln \rho(\omega,\epsilon_i) \right],
\end{equation}
where $P^W((\epsilon_i)$ is the probability distribution of the random on-site energies and
\begin{equation*}
\label{equ:rho_omega_epsilon}
\rho^W(\omega,\epsilon_i) =-\frac{1}{\pi} \mathrm{Im}\, G^W(\omega,\epsilon_i)
\end{equation*}
is the corresponding local density of states.

\item
Obtain the typical Green's function from the TDOS via the Kramers--Kronig relation,
\begin{equation}
\label{equ:typ_green}
G_{\mathrm{typ}}^W(\omega) = \int d\omega' \,
\frac{\rho_{\mathrm{typ}}(\omega')}{\omega^{+}-\omega'}.
\end{equation}

\item
Update the self-energy according to the Dyson equation,
\begin{equation*}
\label{equ:self-energy}
\Sigma^W(\omega) = \omega - \Delta^W(\omega) - \left[G_{\mathrm{typ}}^W(\omega)\right]^{-1} \,.
\end{equation*}

\item
Repeat steps 2--6 until self-consistency is achieved, i.e., until the self-energy (or, equivalently, the TDOS) converges within the desired numerical accuracy.

\end{enumerate}

It is worth noting that replacing the geometric average in Eq.~\eqref{equ:avg} by the arithmetic average reduces the TMT self-consistency loop to the conventional CPA algorithm.

\subsection{The TMT formalism for the Bethe lattice with Cauchy disorder}
\label{sec:tmt_formalism_bethe}

As discussed in the Introduction, we consider a disordered Bethe lattice with Cauchy-distributed on-site disorder,
\begin{equation}
\label{equ:chauchy}
P^W(\epsilon_i) = \frac{1}{\pi} \frac{W}{\epsilon_i^2+W^2},
\end{equation}
where $W$ denotes the disorder strength. In this case, the general TMT self-consistency loop simplifies considerably for two reasons. First, the geometric average entering Eq.~\eqref{equ:avg} can be evaluated analytically for the Cauchy distribution. Second, the Bethe lattice possesses a particularly simple self-consistency relation between the hybridization function and the local Green's function. As a consequence, the Hilbert transform required for a general lattice is replaced by an algebraic relation. Evaluating Eq.~\eqref{equ:avg} analytically yields the following expression for the TDOS (see Appendix~\ref{sec:app:tdos_cauchy} for details):
\begin{equation}
\label{equ:im_g_typ}
\rho_{\mathrm{typ}}^W(\omega) = -\frac{1}{\pi}
\frac{\mathrm{Im}\,\Delta^W(\omega)}{\left(\omega-\mathrm{Re}\,\Delta^W(\omega)\right)^2+\left(W-\mathrm{Im}\,\Delta^W(\omega)\right)^2}.
\end{equation}

The resulting TMT self-consistency loop for the Bethe lattice with Cauchy disorder then consists of the following steps:

\begin{enumerate}

\item
Start from an initial guess for the self-energy, typically $\Sigma^W(\omega)=0$.

\item
Determine the hybridization function using the Bethe-lattice (Green's functions $G_{\cap}$ and) self-consistency condition
\begin{equation*}
\label{equ:hybridization_bethe}
\Delta^W(\omega) = \frac{D^2}{4} G_{\cap}\!\left(\omega-\Sigma^W(\omega)\right) \,,
\end{equation*}
where $D$ denotes the half-bandwidth and 
\begin{equation}
\label{equ:green_bethe}
G_{\cap}(\omega) = \frac{2}{D^2} \left(\omega-\sqrt{\omega^2-D^2}\right)
\end{equation}
is the non-interacting Green's function of the Bethe lattice in the limit of infinite connectivity~\cite{ge.ko.96,ec.ko.05,ko.ec.05}.

\item
Calculate the typical density of states using Eq.~\eqref{equ:im_g_typ}.

\item
Obtain the typical Green's function from the Kramers--Kronig relation, Eq.~\eqref{equ:typ_green}, and update the self-energy according to the Dyson equation,
\begin{equation*}
\label{equ:self-energy_bethe}
\Sigma^W(\omega) = \omega - \frac{D^2}{4} G_{\mathrm{typ}}^W(\omega) - \left[G_{\mathrm{typ}}^W(\omega)\right]^{-1} \,.
\end{equation*}

\item
Repeat steps 2--4 until self-consistency is achieved.

\end{enumerate}

\section{Results}
\label{sec:results}

We consider the Anderson-Hamiltonian, Eq.~\eqref{equ:H_and}, on the infinite-connectivity Bethe lattice with Cauchy-distributed on-site disorder, Eq.~\eqref{equ:chauchy}. Throughout this work, we take the bandwidth $2D$ as the unit of energy.

\subsection{Typical density of states}
\label{sec:tdos_results}

\begin{figure}[h]
    \centering
   \includegraphics[width=0.9\columnwidth]{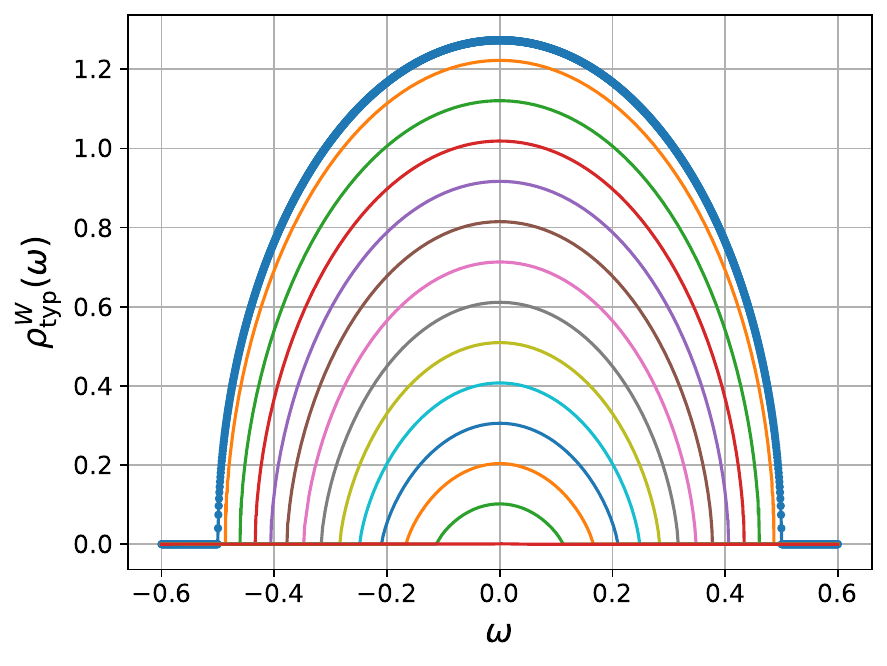}
    \caption{Typical density of states (TDOS) $\rho_{\mathrm{typ}}^W$ on the Bethe lattice for Cauchy-distributed disorder and different disorder strengths $W$, equally spaced between $0.01$ and $0.25$. The blue curve with symbols corresponds to the DOS of the clean Bethe lattice. 
    The half-bandwidth is $D=0.5$.
    }
    \label{fig:bethe_doses}
\end{figure}

The evolution of the TDOS with increasing disorder exhibits two characteristic features. First, the magnitude of the TDOS is continuously suppressed. Second, the energy interval in which the TDOS remains finite progressively shrinks, reflecting the motion of the mobility edge toward the band center. The numerically calculated TDOS for several values of the disorder strength is shown in Fig.~\ref{fig:bethe_doses}. This behavior of TDOS is similar to   the one reported earlier for the box distribution~\cite{do.pa.03}.

\begin{figure}[h]
    \centering
    \includegraphics[width=0.9\columnwidth]{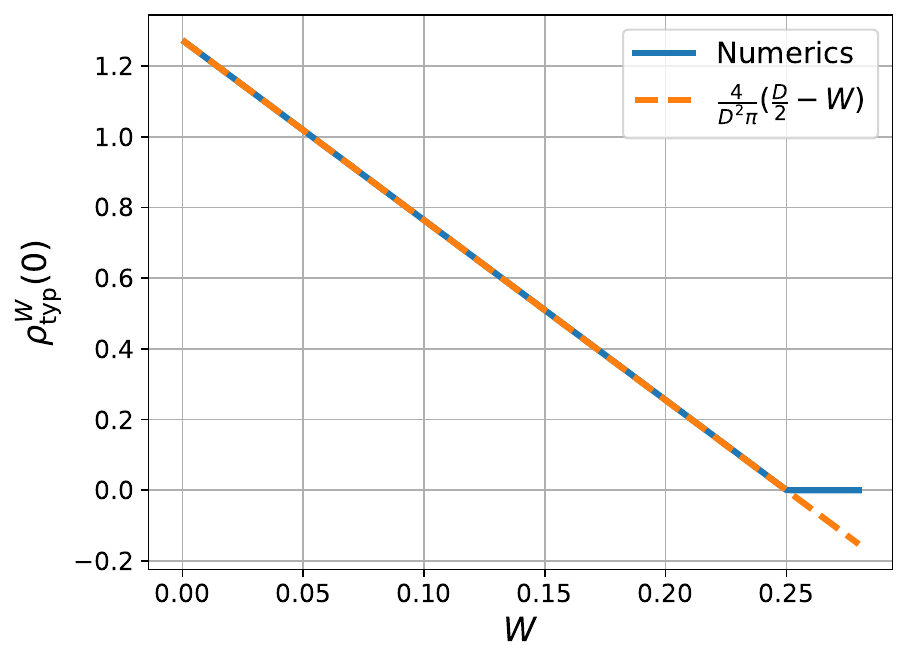}
    \caption{TDOS $\rho_{\mathrm{typ}}^W(\omega)$ at $\omega=0$ as a function of the disorder strength $W$. The orange dashed line shows the analytical result. 
    }
    \label{fig:bethe_doses_zero}
\end{figure}

Owing to particle-hole symmetry, the TDOS reaches its maximum at $\omega=0$. As shown in Fig.~\ref{fig:bethe_doses_zero}, this maximum decreases linearly with increasing disorder, from
\begin{align*}
\rho_{\mathrm{typ}}^{W=0}(0)=\frac{4}{\pi D^2}
\end{align*}
for the clean system ($W=0$) to zero at the critical disorder strength
\begin{align*}
W_c=\frac{D}{2} \,.
\end{align*}

Remarkably, this behavior is not merely a numerical observation but follows exactly from the analytical solution presented in Appendix~\ref{sec:app:tdos_zero_frequency}. The zero-frequency TDOS is given by
\begin{equation}
\label{equ:htyp_vs_W}
\rho_{\mathrm{typ}}^W(0) = \frac{4}{\pi D^2} \left(\frac{D}{2}-W\right) \,,
\end{equation}
which is valid throughout the metallic phase,
$W<\frac{D}{2}$,
while for stronger disorder the TDOS vanishes identically.

\begin{figure}[h]
    \centering
    \includegraphics[width=0.9\columnwidth]{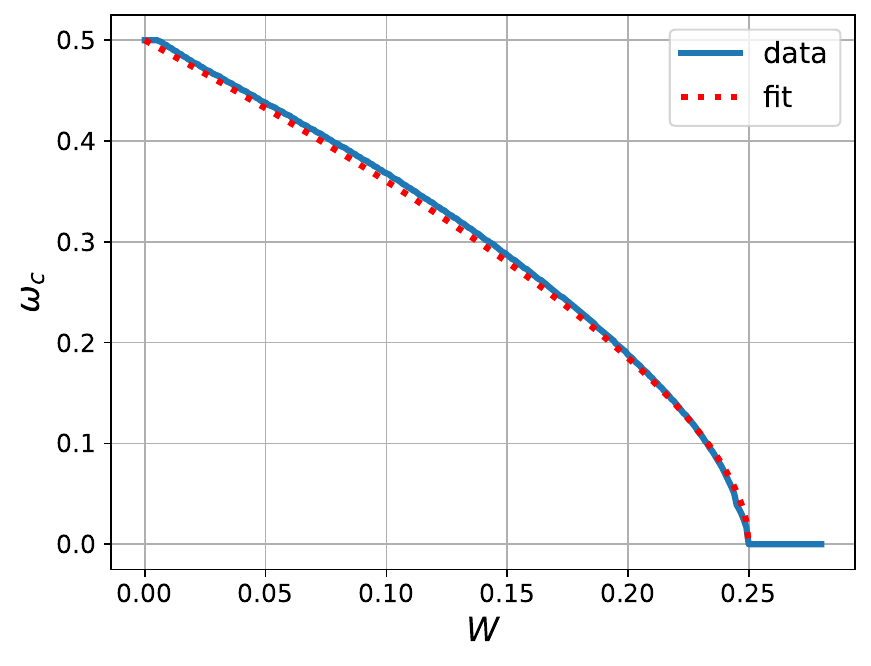}
    \caption{Mobility edge $\omega_c$ as a function of the disorder strength $W$ for the Bethe lattice with Cauchy-distributed disorder. For $|\omega|>\omega_c$, the TDOS vanishes, $\rho_{\mathrm{typ}}^W(\omega)=0$.
    The dotted red line corresponds to a fit obtained from Eq.~\eqref{equ:omegac_fit}, with the fitting parameters $a=0.6874$.
    }
    \label{fig:critical_W}
\end{figure}

The second characteristic effect of increasing disorder is the continuous reduction of the mobility-edge energy. As the disorder strength approaches the critical value $W_c=D/2$, the mobility edge moves toward the band center, leading to a progressive reduction of the energy window occupied by extended states. The numerical results for the mobility edge are presented in Fig.~\ref{fig:critical_W}. In the vicinity of the localization transition, they are well described by the scaling form (see Appendix~\ref{sec:app:analytical_omegac}, where an approximate analytical derivation is presented),
\begin{equation}
\label{equ:omegac_fit}
\omega_c =a D\left(1-\frac{2W}{D}\right)^{1/2} + (1-a) D\left(1-\frac{2W}{D}\right) \,,
\end{equation}
where $a$ is the fitting parameter.

\begin{figure}[h]
    \centering
    \includegraphics[width=0.9\columnwidth]{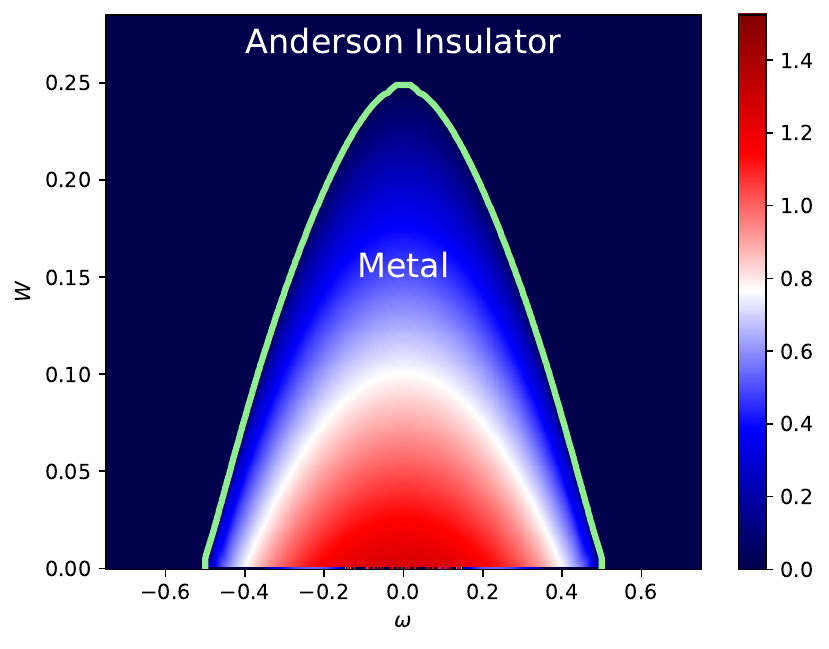}
    \caption{$\omega$--$W$ phase diagram for the Bethe lattice with Cauchy-distributed disorder. The green curve denotes the mobility edge separating the metallic and localized (insulating) phases. The color map shows the TDOS $\rho_{\mathrm{typ}}^W(\omega)$.}
    \label{fig:bethe_phase_diag}
\end{figure}

The combined evolution of the TDOS with energy and disorder is summarized in the $\omega$--$W$ phase diagram shown in Fig.~\ref{fig:bethe_phase_diag}. The green curve denotes the mobility edge separating the metallic and localized regions of the spectrum, while the color map represents the magnitude of the TDOS, $\rho_{\mathrm{typ}}^W(\omega)$. As the disorder strength approaches $W_c=D/2$, the mobility edge collapses to the band center and the metallic region disappears, leaving the entire spectrum Anderson-localized.

\subsection{Extended and localized spectral weigths}
\label{sec:exatended_localized_states}

The TDOS provides a natural way to distinguish between extended and localized electronic states. Since the TDOS is finite only for extended states, its integral directly measures the number of extended states below the chemical potential. The remaining states are therefore localized.

Because disorder redistributes the spectral weight without changing the total number of states, the total number of states is given by
\begin{equation}
\label{equ:Ntot}
N_{\mathrm{tot}} = \int_{-\infty}^{\mu} d\omega\, \rho_{\cap}(\omega) = -\frac{1}{\pi} \int_{-\infty}^{\mu} d\omega\, \mathrm{Im}\, G_{\cap}(\omega)\,,
\end{equation}
where $\mu$ is the chemical potential corresponding to the chosen filling. The number of extended states is defined as
\begin{equation}
\label{equ:Next}
N_{\mathrm{ext}}^W = \int_{-\infty}^{\mu} d\omega\, \rho_{\mathrm{typ}}^W(\omega) \,, 
\end{equation} 
while the number of localized states is obtained from
\begin{equation}
\label{equ:Nloc}
N_{\mathrm{loc}}^W = N_{\mathrm{tot}} - N_{\mathrm{ext}}^W \,. 
\end{equation}

\begin{figure}[h]
    \centering
    \includegraphics[width=0.9\columnwidth]{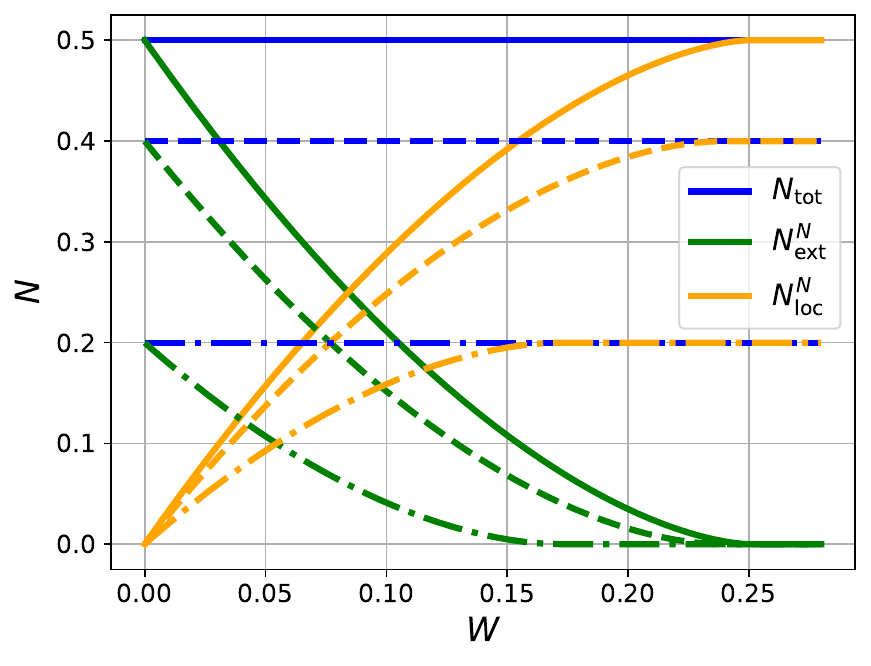}
    \caption{Number of extended and localized states, $N_{\mathrm{ext}}^W$ and $N_{\mathrm{loc}}^W$, as a function of the disorder strength $W$ for the Bethe lattice with Cauchy-distributed disorder. 
    Results are shown at three fillings: half-filling (solid lines), $N_{\mathrm{tot}}=0.4$ (dashed lines), and $N_{\mathrm{tot}}=0.2$ (dash-dotted lines). The color coding of the curves is given in the legend.}
    \label{fig:Next_Nloc_Ntot}
\end{figure}

The disorder dependence of $N_{\mathrm{ext}}^W$ and $N_{\mathrm{loc}}^W$ for several band fillings is shown in Fig.~\ref{fig:Next_Nloc_Ntot}. We first consider the half-filled case. In the absence of disorder, all states are extended, i.e.,
\begin{align*}
N_{\mathrm{ext}}^W=N_{\mathrm{tot}}, \qquad N_{\mathrm{loc}}^W=0 \,.
\end{align*}
As the disorder strength increases, the number of extended states decreases continuously, while the number of localized states increases accordingly. At $W\simeq0.0842$, the two become equal,
$ N_{\mathrm{ext}}^W = N_{\mathrm{loc}}^W =0.25$. With further increasing disorder, localized states become dominant, and at the critical disorder
\begin{align*}
W_c=\frac{D}{2}=0.25,
\end{align*}
all states are localized, i.e.,
\begin{align*}
N_{\mathrm{ext}}^W=0, \qquad N_{\mathrm{loc}}^W=N_{\mathrm{tot}} \,.
\end{align*}

The same qualitative behavior is observed away from half-filling. However, as the filling moves away from half-filling, both the crossover point at which  $N_{\mathrm{ext}}^W=N_{\mathrm{loc}}^W$ and the critical disorder at which all states become localized shift to lower values of $W$. This behavior is fully consistent with the phase diagram shown in Fig.~\ref{fig:bethe_phase_diag}, where the mobility edge reaches the largest disorder strength at the band center ($\omega=0$) and moves toward smaller disorder strengths with increasing distance from the band center. Consequently, states close to $\omega=0$ remain extended over the widest disorder range, whereas states near the band edges localize at substantially weaker disorder.

\begin{figure*}[t!]
\centering
\includegraphics[width=0.9\textwidth]{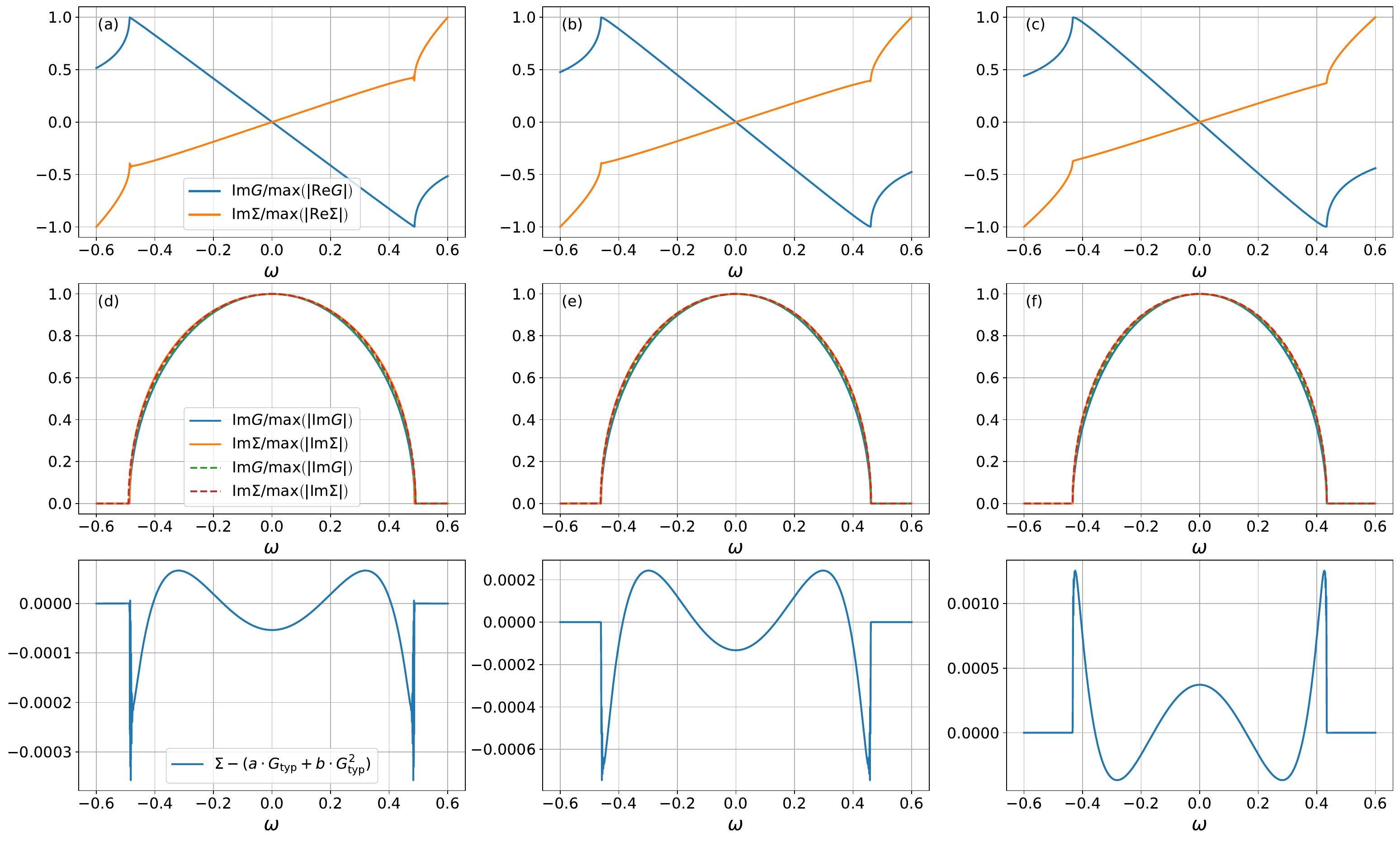}
\caption{The typical Green's function and the self-energy for different disorder strengths. From left to right, the columns correspond to $W=0.01$ [(a,d,g)], $0.03$ [(b,e,h)], and $0.05$ [(c,f,i)]. The first row [(a--c)] shows the normalized real parts, $\mathrm{Re}G_{\mathrm{typ}}^W(\omega)/\max|\mathrm{Re}G_{\mathrm{typ}}^W(\omega)|$ and $\mathrm{Re}\Sigma^W(\omega)/\max|\mathrm{Re}\Sigma^W(\omega)|$, as functions of $\omega$. The second row [(d--f)] shows the normalized imaginary parts, $\mathrm{Im}G_{\mathrm{typ}}^W(\omega)/\max[\mathrm{Im}G_{\mathrm{typ}}^W(\omega)]$ and $\mathrm{Im}\Sigma^W(\omega)/\max[\mathrm{Im}\Sigma^W(\omega)]$, as functions of $\omega$. 
Solid lines show the numerical TMT results, while dashed lines represent the analytical approximation, Eqs.~\eqref{equ:gtyp_sigma_approximate_Final}.
The third row [(g--i)] shows the deviation of the imaginary part of the self-energy from the quadratic fit, $\mathrm{Im}\Sigma-\left(aD^2\,\mathrm{Im}G_{\mathrm{typ}}+bD^3\left(\mathrm{Im}G_{\mathrm{typ}}\right)^2\right)$. 
The fitted coefficients are: for $W=0.01$, $a=0.0982$ and $b=0.0140$; for $W=0.03$, $a=0.3359$ and $b=0.0515$; and for $W=0.05$, $a=0.6530$ and $b=0.1152$.
}
\label{fig:sigma_vs_gtyp}
\end{figure*}

\section{Analytical insights into the metallic TMT solution}
\label{sec:analytical_analysis}

In the previous section we presented a numerical analysis of the TMT solution for the Bethe lattice with Cauchy disorder. The numerical results reveal several remarkably simple features in the metallic regime, suggesting that the essential behavior near the localization transition can be captured by an analytical description. In particular, the typical Green's function and the self-energy exhibit simple scaling properties that allow us to derive approximate analytical expressions in the weak-disorder regime.

The numerical results presented in Fig.~\ref{fig:sigma_vs_gtyp} reveal a remarkably simple structure of the typical Green's function and the self-energy in the metallic phase. As long as the imaginary part of the typical Green's function remains finite, the real parts of both the typical Green's function and the self-energy are accurately described by linear functions of frequency. In addition, the imaginary part of the self-energy is well approximated by a low-order polynomial of the imaginary part of the typical Green's function. These observations provide the basis for the analytical description developed below.

For the Bethe lattice, the typical Green's function is related to the non-interacting Bethe lattice Green's function, Eq.~\eqref{equ:green_bethe}, through the self-energy,
\begin{equation}
\label{equ:gtyp_gbethe}
G_{\mathrm{typ}}^W(\omega) = G_{\cap} \!\left(\omega-\Sigma^W(\omega)\right) \,.
\end{equation}

In the weak-disorder regime the self-energy remains small compared with the bandwidth,
$|\Sigma|\ll D$.
Expanding the Bethe lattice Green's function to linear order in the self-energy yields (see Appendix~\ref{sec:app:analytical} for the derivation)
\begin{align}
\label{equ:G_approximation}
G_{\mathrm{typ}}^W(\omega) \simeq 
\begin{cases}
\dfrac{2}{D^2} \left(\omega-\Sigma^W(\omega) -i\sqrt{-X}\right) \,, &  X<0\,,
\\[10pt]
\dfrac{2}{D^2} \left(\omega-\Sigma^W(\omega) -\mathrm{sgn}(\omega)\sqrt{X} \right)\,, & X>0\,,
\end{cases}
\end{align}
where
\begin{equation}
\label{equ:X_approximation}
X \simeq \omega^2 - D^2 - 2\omega\,\mathrm{Re}\Sigma^W(\omega) \,.
\end{equation}

The numerical results shown in Fig.~\ref{fig:sigma_vs_gtyp} further demonstrate that throughout the metallic regime
\begin{subequations}
\begin{align}
\label{equ:sigma_re_approximate}
\mathrm{Re}\Sigma^W(\omega) &= \alpha\omega \,, 
\\
\label{equ:sigma_im_approximate}
\mathrm{Im}\Sigma^W(\omega) &= aD^2\mathrm{Im}G_{\mathrm{typ}}^W(\omega) + bD^3\left(\mathrm{Im}G_{\mathrm{typ}}^W(\omega)\right)^{\,2} \,,
\end{align}
\end{subequations}
where $a$ and $b$ are fitting parameters with $b\ll1$. The linear frequency dependence of the real part is observed only within the metallic region, where $\mathrm{Im}\,G_{\mathrm{typ}}\neq0$, and provides the basis for the following analytical treatment.

Substituting Eqs.~\eqref{equ:sigma_re_approximate} and \eqref{equ:sigma_im_approximate} into Eq.~\eqref{equ:G_approximation}, and retaining terms up to first order in $b$, yields
\begin{subequations}
\label{equ:gtyp_sigma_approximate_Final}
\begin{align}
\label{equ:gtyp_approximate_Final}
\mathrm{Im}G_{\mathrm{typ}}^W(\omega)
&\simeq
-\frac{1}{\left(a+\frac12\right)D}
\sqrt{1-\frac{\omega^2}{\omega_c^2}}
\nonumber\\
&
\quad
-
\frac{b}{\left(a+\frac12\right)^3D}
\left(
1-\frac{\omega^2}{\omega_c^2}
\right).
\end{align}

Remarkably, the leading contribution retains exactly the square-root form of the Green's function of the clean Bethe lattice. Disorder modifies only the overall amplitude and replaces the non-interacting band edge by the mobility edge $\omega_c$, while the second term represents the leading correction originating from the weak nonlinear dependence of the self-energy on the Green's function.

Using Eq.~\eqref{equ:gtyp_approximate_Final} in Eq.~\eqref{equ:sigma_im_approximate}, again retaining terms up to first order in $b$, gives
\begin{align}
\label{equ::sigma_approximate_Final}
\mathrm{Im}\Sigma^W(\omega)
&\simeq
-\frac{aD}{a+\frac12}
\sqrt{1-\frac{\omega^2}{\omega_c^2}}
\nonumber\\
&
\quad
+
\frac{bD}
{2\left(a+\frac12\right)^3}
\left(
1-\frac{\omega^2}{\omega_c^2}
\right).
\end{align}
\end{subequations}

Thus, both the typical Green's function and the self-energy exhibit the same universal square-root dependence on frequency in the vicinity of the localization transition, while disorder enters only through the renormalized parameters $a$, $b$, and $\omega_c$. This scaling behavior is in excellent agreement with the numerical TMT results presented in Fig.~\ref{fig:sigma_vs_gtyp}.

\section{Conclusion}
\label{sec:conclusion}

In this work we investigated properties of the local density of states on the infinite-connectivity Bethe lattice with Cauchy-distributed on-site disorder within the framework of the typical medium theory (TMT). 
While the corresponding coherent potential approximation (CPA) is also exactly solvable for this disorder distribution, it is unable to describe the Anderson localization transition because it is formulated in terms of the disorder-averaged density of states. In contrast, TMT employs the typical density of states (TDOS) as its central quantity, allowing one to distinguish between extended and localized states through finite and vanishing spectral weights of the TDOS, respectively.

Exploiting the special properties of the Cauchy distribution and the Bethe lattice, we showed that the TMT self-consistency equations simplify considerably and admit several analytical results. In particular, we derived a closed analytical expression for the TDOS and demonstrated that its maximum at the band center decreases linearly with disorder, vanishing at the critical disorder strength $W_c=D/2$. 
While the linear critical behavior is consistent with the general predictions of TMT~\cite{do.pa.03}, the present model provides one of the few cases where these results can be obtained in closed form.
The new results presented here are the analytical determination of $W_c$ and of the mobility edge.
Furthermore, we constructed the $\omega$--$W$ phase diagram and showed that the mobility edge continuously approaches the band center with increasing disorder, leading to complete Anderson localization for $W\geq D/2$.
To further characterize the behavior of the TDOS we investigated the evolution of the numbers of extended and localized states as functions of disorder strength. We found that increasing disorder continuously transfers spectral weight from extended to localized states until the entire spectrum becomes localized at the critical disorder. 

In summary, our analytical and numerical results show that the Lloyd model on the Bethe lattice constitutes an analytically tractable realization of typical medium theory. The special properties of the Cauchy disorder distribution allows naturally to characterize several central quantities of TMT, including the TDOS, the critical disorder, and the mobility-edge evolution.
We believe our results offer a useful reference for extensions that incorporate nonlocal correlations and localization effects beyond the single-site approximation.

\section*{Aknowledgement}
Financial support by the Deutsche Forschungsgemeinschaft (DFG, German Research Foundation) – TRR 360, project no. 492547816, subprojects A5 and C4 –  is gratefully appreciated. 
We have benefited from discussions with D. Vollhardt, V. Dobrosavljevic, and K. Byczuck.  
L.C. gratefully acknowledge the hospitality of University of Oxford, Clarendon Laboratory, UK. \\

\appendix

\section{Analytical derivation of the typical density of states for Cauchy disorder}
\label{sec:app:tdos_cauchy}

In this appendix, we derive the analytical expression for the TDOS for the Cauchy disorder distribution introduced in Sec.~\ref{sec:tmt_formalism_bethe},
\begin{equation}
\label{equ:app:Cauchy}
P^W(\omega)(\epsilon) = \frac{1}{\pi} \frac{W}{\epsilon^2+W^2} \,,
\end{equation}
where $W$ denotes the disorder strength.

The TDOS is defined as the geometric average of the local density of states,
\begin{equation}
\label{equ:app:avg}
\rho_{\mathrm{typ}}^W(\omega) = \exp\left(\int d\epsilon\, P^W(\epsilon)\ln \rho^W(\omega,\epsilon)\right) \,,
\end{equation}
where
\begin{align*}
\rho^W(\omega,\epsilon) &= -\frac{1}{\pi}\mathrm{Im}\,G^W(\omega,\epsilon)
= -\frac{1}{\pi} \mathrm{Im} \frac{1}{\omega^+-\epsilon-\Delta^W(\omega)} \,.
\end{align*}

Using Eqs.~\eqref{equ:app:Cauchy} and \eqref{equ:app:avg}, we obtain
\begin{align*}
\ln \rho_{\mathrm{typ}}^W(\omega) = \int_{-\infty}^{\infty} d\epsilon\, \frac{1}{\pi} \frac{W}{\epsilon^2+W^2}
\ln\left(-\frac{1}{\pi} \mathrm{Im} \frac{1}{\omega-\epsilon-\Delta^W(\omega)} \right) \,.
\end{align*}

We decompose the hybridization function as
\begin{align*}
\Delta^W(\omega)=\mathrm{Re}\Delta^W(\omega) +i \mathrm{Im}\Delta^W(\omega) = \Delta_R+i\Delta_I \,.
\end{align*}
Since $\Delta^W(\omega)$ is the retarded hybridization function, its imaginary part satisfies
$\Delta_I\leq0$ .

For brevity, we suppress the frequency dependence of $\Delta_R$ and $\Delta_I$ in the following derivation. After straightforward algebra, we obtain

\begin{widetext}
\begin{align}
\label{equ:app:rho_typ}
\ln \rho_{\mathrm{typ}}^W(\omega)
&= \int_{-\infty}^{\infty} d\epsilon\, \frac{W}{\pi(\epsilon^2+W^2)}
\ln\left[-\frac{1}{\pi} \frac{\Delta_I}{(\omega-\epsilon-\Delta_R)^2+\Delta_I^2}\right]
\nonumber\\
&
= \int_{-\infty}^{\infty} d\epsilon\, \frac{W}{\pi} \frac{\ln\left(-\Delta_I/\pi\right)}{\epsilon^2+W^2}
- \int_{-\infty}^{\infty} d\epsilon\, \frac{W}{\pi}
\frac{\ln\left((\omega-\epsilon-\Delta_R)^2+\Delta_I^2\right)}{\epsilon^2+W^2}
\nonumber\\
&
=\ln\left(-\frac{\Delta_I}{\pi}\right)
- \int_{-\infty}^{\infty} d\epsilon\, \frac{W}{\pi}
\frac{\ln\left((\omega-\epsilon-\Delta_R)^2+\Delta_I^2\right)}{\epsilon^2+W^2} \,.
\end{align}

In obtaining the second line of Eq.~\eqref{equ:app:rho_typ}, we used that
$\ln(-\Delta_I/\pi)$ is independent of the integration variable $\epsilon$.
Therefore, the remaining integral over the Cauchy distribution is normalized to unity.

The second integral in Eq.~\eqref{equ:app:rho_typ} can be evaluated analytically. Integrals of this type also appear in the computation of the cross-entropy between two Cauchy distributions and were considered by Chyzak and Nielsen~\cite{ch.ni.19}. Specifically, Eq.~(9) of Ref.~\cite{ch.ni.19} gives the following expression:
\begin{equation}
A(a,b,c;d,e,f) = \int_{-\infty}^{\infty} dx\, \frac{\ln(dx^2+ex+f)}{ax^2+bx+c}
=\frac{2\pi \left[\ln \left(2af-be+2cd +\sqrt{4ac-b^2}\sqrt{4df-e^2} \right)-\ln(2)\right]} {\sqrt{4ac-b^2}} \,.
\end{equation}
\end{widetext}

The above expression is valid provided
\begin{align*}
4ac>b^2\,, \qquad 4df>e^2 \,,
\end{align*}
which ensures that both quadratic polynomials are positive on the real axis. In our case, the parameters are
\begin{align*}
a=1,\qquad b=0,\qquad c=W^2 \,,
\end{align*}
and
\begin{align*}
d=1,\qquad e=-2(\omega-\Delta_R), \qquad f=(\omega-\Delta_R)^2+\Delta_I^2 \,.
\end{align*}
Therefore,
\begin{align*}
4ac-b^2=4W^2>0 \,,
\end{align*}
and
\begin{align*}
4df-e^2 &= 4\left[(\omega-\Delta_R)^2+\Delta_I^2\right] - 4(\omega-\Delta_R)^2
\\
&= 4\Delta_I^2>0 \,.
\end{align*}
Consequently,
\begin{align*}
\sqrt{4ac-b^2}=2W \,,
\quad\mathrm{and}\quad
\sqrt{4df-e^2} = 2|\Delta_I| = -2\Delta_I \,,
\end{align*}
where in the last equality we used $\Delta_I\leq0$ for the retarded hybridization function.

Substituting these expressions into the analytical result above and comparing with Eq.~\eqref{equ:app:rho_typ}, we obtain
\begin{align}
\ln \rho_{\mathrm{typ}} &= \ln\left(-\frac{\Delta_I}{\pi}\right) - \ln\left((\omega-\Delta_R)^2+(W-\Delta_I)^2 \right)
\nonumber\\
&=\ln\left(-\frac{1}{\pi} \frac{\Delta_I} {(\omega-\Delta_R)^2+(W-\Delta_I)^2} \right) \,.
\end{align}

Finally, exponentiating the above expression gives the analytical result for the TDOS,
\begin{equation}
\label{equ:app:rho_typ_final}
\rho_{\mathrm{typ}}^W(\omega) = -\frac{1}{\pi} \frac{\Delta_I} {(\omega-\Delta_R)^2+(W-\Delta_I)^2} \,.
\end{equation}

\section{Zero-frequency TDOS for the Bethe lattice with Cauchy disorder}
\label{sec:app:tdos_zero_frequency}

In this appendix, we derive an analytical expression for the maximal value of the TDOS for the Bethe lattice with Cauchy disorder. Due to particle-hole symmetry, the maximum of the TDOS is located at $\omega=0$. We start from Eq.~\eqref{equ:im_g_typ} and use the Bethe lattice self-consistency relation between the typical Green's function and the hybridization function,
\begin{equation}
\label{equ:app:green_vs_hybrid}
G_{\mathrm{typ}}^W(\omega) = \frac{4}{D^2}\Delta^W(\omega) \,.
\end{equation}

Combining Eq.~\eqref{equ:app:green_vs_hybrid} with Eq.~\eqref{equ:im_g_typ}, and using
\begin{equation}
\label{equ:app:rho_vs_green}  
\rho_{\mathrm{typ}}^W(\omega) = -\frac{1}{\pi}\mathrm{Im}G_{\mathrm{typ}}^W(\omega) \,,
\end{equation}
we obtain
\begin{align}
\label{equ:app:condition}
\left(\omega-\mathrm{Re}\Delta^W(\omega)\right)^2 + \left(W-\mathrm{Im}\Delta^W(\omega)\right)^2 = \frac{D^2}{4},
\end{align}
for the metallic solution with nonzero hybridization function.

At zero frequency, particle-hole symmetry implies
\begin{align*}
\mathrm{Re}\Delta^W(0)=0.
\end{align*}
Therefore,
\begin{equation}
\label{equ:app:Delta-W}
\mathrm{Im}\Delta^W(0)-W  = \mp\frac{D}{2}.
\end{equation}

Since $\Delta^W(\omega)$ is a retarded hybridization function, its imaginary part satisfies
\begin{align*}
\mathrm{Im}\Delta^W(\omega)\leq0.
\end{align*}
Consequently, the physically relevant solution in the metallic phase is
\begin{equation}
\label{equ:app:Delta_vs_W}
\mathrm{Im}\Delta^W(0) = -\frac{D}{2}+W .
\end{equation}

Using Eqs.~\eqref{equ:app:green_vs_hybrid} and \eqref{equ:app:rho_vs_green} we finally obtain
\begin{equation}
\label{equ:app:rhtyp_vs_W}
\rho_{\mathrm{typ}}^W(0) = \frac{4}{\pi D^2} \left(\frac{D}{2}-W\right).
\end{equation}

This expression is valid for the metallic phase
$W<\frac{D}{2}$,
while for larger disorder strengths the TDOS vanishes.

\section{Approximate analytical expressions for the Green's function and self-energy}
\label{sec:app:analytical}

In this appendix we derive approximate analytical expressions for the typical Green's function and the self-energy in the weak-disorder regime. The derivation is motivated by the numerical results presented in Fig.~\ref{fig:sigma_vs_gtyp}. As demonstrated there, in the metallic phase, where the imaginary part of the typical Green's function remains finite, the real parts of both the typical Green's function and the self-energy are accurately described by linear functions of frequency. Furthermore, the imaginary part of the self-energy can be represented by a low-order polynomial in the imaginary part of the typical Green's function. These numerical observations provide the basis for the analytical approximations derived below.

For the Bethe lattice, the typical Green's function is related to the non-interacting Bethe lattice Green's function through the self-energy,
\begin{equation}
G_{\mathrm{typ}}^W(\omega) = G_{\cap}(\omega-\Sigma^W(\omega)) \,. \nonumber 
\end{equation}
Using the analytical expression for the Bethe lattice Green's function, we obtain
\begin{align*}
G_{\mathrm{typ}}^W(\omega) = \frac{2}{D^2} \left(\omega-\Sigma^W(\omega) - \sqrt{\left(\omega-\Sigma^W(\omega)\right)^2-D^2} \right) \,.
\end{align*}

For simplicity, in the following we suppress the explicit frequency dependence of all quantities. We denote the real and imaginary parts of the Green's function and the self-energy by the subscripts $R$ and $I$, respectively. Thus,
\begin{align}
G_{\mathrm{typ}} &= \frac{2}{D^2} \left(\omega-\Sigma - \sqrt{\omega^2-2\omega\Sigma+\Sigma^2-D^2} \right)
\nonumber\\
&=\frac{2}{D^2} \Bigl(\omega-\Sigma_R-i\Sigma_I 
\nonumber\\
&\qquad - \sqrt{(\omega-\Sigma_R)^2-\Sigma_I^2 - D^2-2i(\omega-\Sigma_R)\Sigma_I} \Bigr)\,.
\nonumber
\end{align}

Introducing
\begin{subequations}
\begin{align}
X&= (\omega-\Sigma_R)^2-\Sigma_I^2-D^2 \,, \\
Y&= -2(\omega-\Sigma_R)\Sigma_I \,,
\end{align}
\end{subequations}
the Green's function can be written as
\begin{align*}
G_{\mathrm{typ}} = \frac{2}{D^2} \left(\omega-\Sigma_R-i\Sigma_I-\sqrt{X+iY} \right) \,.
\end{align*}

Using the standard decomposition of a complex square root into its real and imaginary parts, one obtains
\begin{subequations}
\begin{align}
G_R &= \frac{2}{D^2} \left[\omega-\Sigma_R - \mathrm{sgn}(\omega) \sqrt{\frac{\sqrt{X^2+Y^2}+X}{2}} \right] \,,
\\
G_I &= \frac{2}{D^2} \left[-\Sigma_I - \sqrt{\frac{\sqrt{X^2+Y^2}-X}{2}} \right] \,.
\end{align}
\end{subequations}

We now consider the weak-disorder regime, where the self-energy is small compared with the bandwidth,
$|\Sigma_R|,|\Sigma_I|\ll D$.
Keeping only terms linear in the self-energy gives
\begin{equation}
\label{equ:app:X_approximation}
X\simeq \omega^2-D^2-2\omega\Sigma_R , \qquad Y\simeq -2\omega\Sigma_I \,.
\end{equation}

Since $Y$ is already first order in the self-energy, the contribution $Y^2$ appears only in second order and can be neglected. Consequently, the real and imaginary parts of the typical Green's function become
\begin{subequations}
\label{equ:app:G_approximation}
\begin{align}
G_{\mathrm{typ},R} &\simeq 
\begin{cases} 
\dfrac{2}{D^2} (\omega-\Sigma_R), & X<0 \,, 
\\[10pt]
\dfrac{2}{D^2} \left(\omega-\Sigma_R-\mathrm{sgn}(\omega)\sqrt{X} \right), & X>0 \,,
\end{cases}
\\[8pt]
G_{\mathrm{typ},I} &\simeq
\begin{cases}
\dfrac{2}{D^2} \left(-\Sigma_I-\sqrt{-X}\right), & X<0 \,,
\\[10pt]
-\dfrac{2}{D^2}\Sigma_I, & X>0 \,.
\end{cases}
\end{align}
\end{subequations}
We note that $\Sigma_I$ vanishes in the localized phase, corresponding to $X>0$.

The numerical results shown in Fig.~\ref{fig:sigma_vs_gtyp} demonstrate that in the metallic phase, characterized by a finite imaginary part of the typical Green's function, the real parts of both the typical Green's function and the self-energy exhibit a linear frequency dependence. Therefore, we introduce the following approximation,
\begin{subequations}
\label{equ:app:selfenergy_vs_gtyp}
\begin{equation}
\label{equ:app:selfenergy_vs_gtyp_R}
G_{\mathrm{typ},R} = \frac{\alpha'}{D^2}\omega \,,
\qquad
\Sigma_R =\alpha\omega \,,
\end{equation}
where $\alpha'>0$ and $\alpha<0$.

The linear frequency dependence observed in Fig.~\ref{fig:sigma_vs_gtyp} characterizes the metallic regime, where the imaginary part of the typical Green's function is finite. Motivated by this behavior, we retain the leading linear contribution of the real part of the self-energy in the following analysis. Outside the mobility edge, where the TDOS vanishes, the frequency dependence of the self-energy changes and the linear behavior is no longer observed. Since the analytical derivation below is focused on the metallic solution and its evolution towards the localization transition, we do not consider the detailed behavior of the self-energy in the localized regime. Importantly, this behavior has no influence on the results and conclusions presented in this paper.

The numerical results further show that, within the metallic phase, the imaginary part of the self-energy is accurately described by a quadratic function of the imaginary part of the typical Green's function,
\begin{equation}
\label{equ:app:selfenergy_vs_gtyp_I}
\Sigma_I \simeq aD^2G_{\mathrm{typ},I} + bD^3G_{\mathrm{typ},I}^{2} \,,
\end{equation}
where $a$ and $b$ are fitting parameters.
\end{subequations}

We now focus on the metallic regime, where $G_{\mathrm{typ},I}\neq0$ and consequently $X<0$. Substituting the linear form of the real part of the self-energy, Eq.~\eqref{equ:app:selfenergy_vs_gtyp_R}, into Eq.~\eqref{equ:app:X_approximation}, we obtain
\begin{align}
\label{equ:app:X_approximate_final}
X &\simeq \omega^2-D^2-2\alpha\omega^2  = (1-2\alpha)\omega^2-D^2
\nonumber\\
&= -\frac{D^2}{\omega_c^2} \left(\omega_c^2-\omega^2\right)\,,
\end{align}
where we have introduced the characteristic energy scale
\begin{equation}
\omega_c
=
\frac{D}{\sqrt{1-2\alpha}} .
\end{equation}

The parameter $\omega_c$ naturally appears as the boundary between the regions with $X<0$ and $X>0$ and therefore corresponds to the mobility edge obtained within this approximation.

Next, substituting Eqs.~\eqref{equ:app:selfenergy_vs_gtyp_I} and
\eqref{equ:app:X_approximate_final} into the expression for the imaginary part of the Green's function, Eq.~\eqref{equ:app:G_approximation}, gives
\begin{align*}
G_{\mathrm{typ},I} &= \frac{2}{D^2} \left(-aD^2G_{\mathrm{typ},I} -bD^3G_{\mathrm{typ},I}^{2} -\sqrt{-X} \right) \,.
\end{align*}

After rearranging terms, this becomes a quadratic equation for
$G_{\mathrm{typ},I}$,
\begin{align*}
bD^3G_{\mathrm{typ},I}^{2} + D^2\left(a+\frac12\right)G_{\mathrm{typ},I} + \frac{D}{\omega_c} \sqrt{\omega_c^2-\omega^2} =0 \,.
\end{align*}
Solving this equation yields
\begin{align}
G_{\mathrm{typ},I} = \frac{-D^2\left(a+\frac12\right) \pm \sqrt{D^4\left(a+\frac12\right)^2 - 4bD^4 \sqrt{1 - \frac{\omega_c^2}{\omega_c^2}}}}{2bD^3} \,.
\nonumber
\end{align}
Equivalently,
\begin{align*}
G_{\mathrm{typ},I} = -\frac{a+\frac12}{2bD} \left[ 1 \mp \sqrt{ 1- \frac{4b}{(a+\frac12)^2} \sqrt{1-\frac{\omega^2}{\omega_c^2}}} \right] \,.
\end{align*}

Since the coefficient $b$ is small, we expand the square root to first order in $b$. Selecting the physical branch that remains finite in the limit $b\rightarrow0$, we obtain
\begin{subequations}
\begin{equation}
\label{equ:app:Gtyp_approximate_Final}
G_{\mathrm{typ},I} \simeq -\frac{1}{(a+\frac12)D} \sqrt{1-\frac{\omega^2}{\omega_c^2}} - \frac{b}{(a+\frac12)^3D} \left(1-\frac{\omega^2}{\omega_c^2}\right) \,.
\end{equation}

Finally, we substitute Eq.~\eqref{equ:app:Gtyp_approximate_Final} into the relation between the imaginary parts of the self-energy and the typical Green's function, Eq.~\eqref{equ:app:selfenergy_vs_gtyp_I}. Keeping terms up to first order in $b$, we obtain
\begin{align*}
\Sigma_I &\simeq -\frac{aD}{a+\frac12} \sqrt{1-\frac{\omega^2}{\omega_c^2}} - \frac{abD}{(a+\frac12)^3}\left(1-\frac{\omega^2}{\omega_c^2}\right)
\nonumber\\
&\qquad + \frac{bD}{(a+\frac12)^2}\left(1-\frac{\omega^2}{\omega_c^2}\right)
\nonumber\\
&= -\frac{aD}{a+\frac12} \sqrt{1-\frac{\omega^2}{\omega_c^2}} + \frac{bD}{2(a+\frac12)^3} \left(1-\frac{\omega^2}{\omega_c^2}\right) \,.
\end{align*}

Therefore, the approximate analytical form of the imaginary part of the self-energy in the metallic phase is
\begin{equation}
\label{equ:app:sigma_approximate_Final}
\Sigma_I \simeq -\frac{aD}{a+\frac12} \sqrt{1-\frac{\omega^2}{\omega_c^2}} + \frac{bD}{2(a+\frac12)^3}\left(1-\frac{\omega^2}{\omega_c^2} \right) \,.
\end{equation}
\end{subequations}

\begin{figure}[t]
    \centering
    \includegraphics[width=0.9\columnwidth]{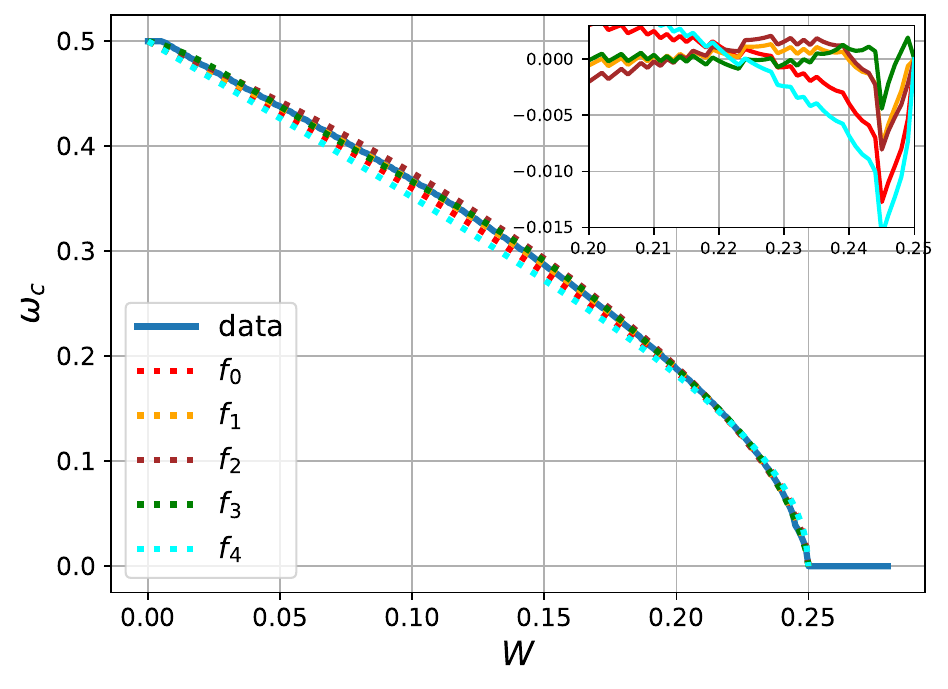}
    \caption{Comparison of different fitting functions for the disorder dependence of the mobility edge $\omega_c(W)$ of the Bethe lattice with Cauchy-distributed disorder. The fits $f_0$--$f_4$ are obtained from Eqs.~\eqref{equ:omegac_fit} and \eqref{equ:app:fit_functions}, respectively. Their corresponding fitting parameters are: 
    $f_0$: $a_0=0.6874$; 
    $f_1$: $p_1=0.6059$;  
    $f_2$: $a_2=0.5560$, $b_2=0.8292$; 
    $f_3$: $a_3=0.4428$, $b_3=-0.9199$, $c_3=32.58$, $d_3=-176.1$, $e_3=445.4$, $g_3=-590.4$, $h_3=395.8$;
    and 
    $f_4$: $a_4=0.7694$;. The inset shows the difference between the numerical data and the corresponding fitted curves.}
    \label{fig:critical_W_extra}
\end{figure}

\section{Analytical derivation of the disorder dependence of the mobility edge}
\label{sec:app:analytical_omegac}

Within TMT, the localization transition is identified by the vanishing of the TDOS,
${\rho_{\mathrm{typ}}=0}$. At zero frequency, our numerical results exhibit the asymptotic behavior
\begin{align*}
\rho_{\mathrm{typ}}^W(0)\propto(W_c-W) \,,
\end{align*}
corresponding to the order-parameter critical exponent $\beta=1$. 
Whether this exponent is a consequence of the TMT approximation, the Cauchy disorder distribution, or the mathematical structure of the Bethe lattice remains an open question.

Nevertheless, the observed critical behavior allows us to discuss the analytical form of the mobility edge close to the localization transition.
In the vicinity of the critical point, the TDOS can be expanded analytically around $\omega=0$ and $W=W_c$ as
\begin{align*}
\rho_{\mathrm{typ}}^W(\omega) &= \sum_{n=1}^{\infty} a_n(W_c-W)^n + \sum_{m=1}^{\infty} b_m\omega^m 
\nonumber\\
&+ \sum_{n,m=1}^{\infty} c_{n,m}(W-W_c)^n\omega^m \,. 
\end{align*}
Introducing $\delta=W_c-W\rightarrow0^+$ \,,
this represents an expansion around the point
$(\omega,\delta)=(0,0)$.
Because of particle-hole symmetry,
$\omega\leftrightarrow-\omega$,
both the non-interacting Bethe-lattice Green's function $G_{\cap}$ and the TMT Green's function $G_{\mathrm{typ}}$ are even functions of frequency.
Consequently, the TDOS inherits the same symmetry and all odd powers of $\omega$ vanish from the expansion:
\begin{align*}
\rho_{\mathrm{typ}}^W(\omega) &= \sum_{n=1}^{\infty} a_n(W_c-W)^n + \sum_{m=1}^{\infty}b_{2m}\omega^{2m}
\nonumber\\
&+ \sum_{n,m=1}^{\infty} c_{n,2m}(W-W_c)^n\omega^{2m} \,.
\end{align*}

The coefficient $a_1$ follows directly from Eq.~\eqref{equ:htyp_vs_W},
yielding: $a_1={4}/{(\pi D^2)}$.
Retaining only the leading contributions and imposing the condition
$\rho_{\mathrm{typ}}=0$
immediately gives
\begin{align}
\label{equ:app:omega_c_analytical}
\omega_c=
\sqrt{\frac{a_1}{b_2}}
(W_c-W)^{1/2}.
\end{align}

Equation~\eqref{equ:app:omega_c_analytical} predicts the square-root critical behavior of the mobility edge in the immediate vicinity of the localization transition.
The fitting expression used in the main text, Eq.~\eqref{equ:omegac_fit}, should therefore be regarded as an empirical interpolation formula that preserves this exact asymptotic behavior while extending the agreement with the numerical results over a much broader disorder range.
Very close to $W_c$, the square-root contribution dominates, whereas the linear correction becomes increasingly important away from the critical point and governs the crossover from the critical regime.

To assess the robustness of the fitting procedure, we also considered several alternative fitting functions:
\begin{widetext}
\begin{subequations}
\label{equ:app:fit_functions}
\begin{align}
\label{equ:app:fit_function_1}
f_1(W) &= D\left(1-\frac{W}{W_c}\right)^{p_1}, 
\\ 
\label{equ:app:fit_function_2}
f_2(W) &= a_2D\left(1-\frac{W}{W_c}\right)^{1/2} + b_2D\left(1-\frac{W}{W_c}\right)
+
c_2D\left(1-\frac{W}{W_c}\right)^{3/2},
\\
\label{equ:app:fit_function_3}
f_3(W) &= a_3D\left(1-\frac{W}{W_c}\right)^{1/2}
+
b_3D\left(1-\frac{W}{W_c}\right)
+
c_3D\left(1-\frac{W}{W_c}\right)^{3/2}
+
d_3D\left(1-\frac{W}{W_c}\right)^2
\nonumber\\
&
+
e_3D\left(1-\frac{W}{W_c}\right)^{5/2}
+
g_3D\left(1-\frac{W}{W_c}\right)^3
+
h_3D\left(1-\frac{W}{W_c}\right)^{7/2}
+
i_3D\left(1-\frac{W}{W_c}\right)^4,
\\
\label{equ:app:fit_function_4}
f_4(W)
&=
a_4D\left(1-\frac{W}{W_c}\right)^{1/2}
+
b_4D\left(1-\frac{W}{W_c}\right)^{3/2}.
\end{align}
\end{subequations}
\end{widetext}
To reduce the number of independent fitting parameters and ensure the correct clean-limit value $\omega_c(0)=D$, we impose the normalization conditions
\begin{align*}
&a_2+b_2+c_2 =1, \\
&a_3+b_3+c_3+d_3+e_3+g_3+h_3+i_3 =1, \\
&a_4+b_4 =1 .
\end{align*}

The corresponding results are presented in Fig.~\ref{fig:critical_W_extra}.
Equation~\eqref{equ:app:fit_function_1} provides a good description of the numerical data over a broad disorder range.
However, the fitted exponent $p$ is larger than the analytically expected value $1/2$.
Such an exponent should therefore be regarded as an effective exponent reflecting the finite fitting interval rather than the true asymptotic critical behavior.
For this reason, we prefer fitting functions whose leading contribution is proportional to $(W_c-W)^{1/2}$.

The fitting function given by Eq.~\eqref{equ:app:fit_function_2} includes the next term in the expansion in powers of $(W_c-W)^{1/2}$ compared with the interpolation formula used in the main text, Eq.~\eqref{equ:omegac_fit}.
As shown in Fig.~\ref{fig:critical_W_extra}, this additional term significantly extends the range over which the analytical expression reproduces the numerical mobility edge.
Including still higher-order terms, as in Eq.~\eqref{equ:app:fit_function_3}, further improves the quality of the fit and reproduces the numerical data over essentially the entire metallic regime.

Finally, we also considered the fitting function given by Eq.~\eqref{equ:app:fit_function_4}.
This expression follows naturally from retaining the $\omega^4$ contribution in the analytical expansion and expressing the result as a series in powers of $(W_c-W)^{1/2}$.
Although analytically motivated, it provides a noticeably poorer description of the numerical data than the other fitting functions.
This indicates that higher-order contributions neglected in the simple expansion become important before the quartic correction alone is sufficient to describe the numerical results.

To conclude, Eq.~\eqref{equ:omegac_fit} represents an empirical interpolation formula that preserves the exact square-root critical behavior close to $W_c$ while providing an accurate description of the numerical mobility edge over a substantially broader disorder range.
Higher-order expansions systematically improve the quality of the fit but do not modify the leading critical behavior.

\bibliography{ref}

\end{document}